\documentclass[11pt]{article}
\pdfoutput=1 

\usepackage{jheppub} 
                     
\usepackage{braket,slashed,bm}
\usepackage{array,multirow}

\usepackage[T1]{fontenc} 

\usepackage{xparse}
\usepackage{etoolbox}

\usepackage{booktabs}
\usepackage{multirow}
\usepackage{footmisc}

\def\nn{\nonumber\\ }
\def\rd{{\rm d}}
\def\abs#1{\left| #1 \right| }
\def\lchi{\Lambda_\chi}

\def\lbsm{\Lambda_{\text{BSM}}}

\def\vev#1{\big\langle #1 \big\rangle}

\renewcommand{\L}{\mathcal{L}}
\renewcommand{\O}{\mathcal{O}}
\newcommand{\TeV}{\,\text{TeV}}
\newcommand{\GeV}{\,\text{GeV}}
\newcommand{\MeV}{\,\text{MeV}}
\newcommand{\hc}{\mathrm{h.c.}}
\newcommand{\p}{\partial}
\NewDocumentCommand{\lwc}{ m m O{} o }{
	L^{\ifblank{#3}{}{#3,}#2 }_{\IfNoValueTF{#4}{#1}{\substack{#1\\#4}}}
}
\NewDocumentCommand{\lwct}{ m m O{} o }{
	\widetilde L^{\ifblank{#3}{}{#3,}#2 }_{\IfNoValueTF{#4}{#1}{\substack{#1\\#4}}}
}
\NewDocumentCommand{\Op}{ m m O{} o }{
	\O^{\ifblank{#3}{}{#3,}#2 }_{\IfNoValueTF{#4}{#1}{\substack{#1\\#4}}}
}

\renewcommand{\O}{\mathcal{O}}
\newcommand{\op}[3]{\O^{#2,#3}_{#1}}

\newcommand{\wc}[3]{L^{#2,#3}_{#1}}

\NewDocumentCommand{\dlwc}{ m m O{} o }{
	{\dot L}^{\ifblank{#3}{}{#3,}#2 }_{\IfNoValueTF{#4}{#1}{\substack{#1\\#4}}}
}

\newcommand{\nnn}{\nonumber\\[-0.4cm] }

\newcommand{\chpt}{$\chi$PT}

\title{\boldmath Non-Perturbative Effects in $\mu\to e\gamma$
}

\author[a]{Wouter Dekens,}

\author[a]{Elizabeth E.~Jenkins,}

\author[a]{Aneesh V.~Manohar,}

\author[a,1]{Peter Stoffer}\note{Corresponding author.}

\affiliation[a]{Department of Physics, University of California at San Diego, 9500 Gilman Drive,\\ La Jolla, CA 92093-0319, USA}

\emailAdd{wdekens@ucsd.edu}
\emailAdd{ejenkins@ucsd.edu}
\emailAdd{amanohar@ucsd.edu}
\emailAdd{pstoffer@ucsd.edu}

\abstract{We compute the non-perturbative contribution of semileptonic tensor operators $(\bar q \sigma^{\mu \nu} q)(\bar \ell \sigma_{\mu \nu} \ell)$ to the purely leptonic process $\mu \to e \gamma$ and to the electric and magnetic dipole moments of charged leptons by matching onto chiral perturbation theory at low energies. 
This matching procedure has been used extensively
to study semileptonic and leptonic weak decays of hadrons.  In this paper, we apply it to observables that contain no strongly interacting external particles. The non-perturbative contribution
to $\mu \to e $ processes is used to extract the best current bound on lepton-flavor-violating semileptonic tensor operators, $\lbsm \gtrsim 450$\,TeV. We briefly discuss how the same method applies to dark-matter interactions.}

\begin{document} 
\maketitle

\section{Introduction}\label{sec:intro}

The branching ratio for the purely leptonic decay $\mu \to e \gamma$ has the current upper bound
\begin{align}
	\label{eq:BRmuegamma}
	\text{BR}(\mu \to e \gamma) < 4.2 \times 10^{-13} \, ,
\end{align}
determined by the MEG collaboration~\cite{Adam:2013mnn,TheMEG:2016wtm}. In the Standard Model (SM) including flavor-changing Majorana neutrino masses, the decay rate is suppressed by $(m_\nu/M_W)^4 \sim 10^{-48}$, so the result Eq.~\eqref{eq:BRmuegamma} provides strong constraints on $\mu \to e$ transitions due to new physics beyond the Standard Model (BSM). The effect of new physics at energies  above the muon mass can be described by an effective field theory, and the leading operators that contribute to $\mu \to e \gamma$ are the dipole operators\footnote{We use the conventions of Refs.~\cite{Jenkins:2017dyc,Jenkins:2017jig}. Other commonly used normalization conventions include factors of $e$ and/or $m_\mu$.}
\begin{align}
	\label{eq:DipoleOperator}
	\mathcal{L} &=
	\lwc{e\gamma}{}[][e\mu] \, \overline e_L \sigma^{\mu \nu} \mu_R  \, F_{\mu \nu} 
	+ \lwc{e\gamma}{}[][\mu e] \, \overline \mu_L \sigma^{\mu \nu} e_R  \, F_{\mu \nu} 
	+ \hc
\end{align}
These dimension-five operators result in a decay width
\begin{align}
	\label{eq:DecayRate}
	\Gamma(\mu \to e \gamma) &= \frac{m_\mu^3}{4 \pi} \left( \big| \lwc{e\gamma}{}[][e\mu] \big|^2 +  \big| \lwc{e\gamma}{}[][\mu e] \big|^2  \right) + \O(m_e^2) \, ,
\end{align}
so Eq.~\eqref{eq:BRmuegamma} places the limit
\begin{align}
	\label{eq:LimitWilsonCoeff}
	\sqrt{ \big| \lwc{e\gamma}{}[][e\mu] \big|^2 +  \big| \lwc{e\gamma}{}[][\mu e] \big|^2 } < 3.7 \times 10^{-11} \TeV^{-1} \, .
\end{align}

If one assumes that new physics consists of particles with masses well above the electro\-weak scale, the Standard Model Effective Field Theory (SMEFT) provides an adequate description of BSM effects~\cite{Buchmuller:1985jz,Grzadkowski:2010es}. At the weak scale, the SMEFT is matched to the low-energy effective field theory (LEFT)~\cite{Jenkins:2017dyc,Jenkins:2017jig}. In such a scenario, the dimension-five dipole operators in Eq.~\eqref{eq:DipoleOperator} derive from dimension-six dipole operators in the SMEFT, so their coefficients are of the same size as those of the other dimension-six operators. 
In this paper, we will consider the case where $L_{e \gamma}$ is formally of order $1/\lbsm^2$, and of the same size as other dimension-six operators.
In many BSM models, the dipole operators are typically induced with a coefficient $\lwc{e\gamma}{} \sim e m_\mu/\lbsm^2$, and so the limit Eq.~\eqref{eq:LimitWilsonCoeff} is equivalent to 
$\lbsm > 930\TeV$.

The operators in the  low-energy effective theory below the weak scale have been classified in Ref.~\cite{Jenkins:2017dyc}. We can calculate the decay rate $\mu \to e \gamma$ in terms of LEFT operators renormalized at a scale $\mu \sim 2\GeV$ at which perturbation theory is still valid. The heavy quarks can be integrated out of the theory, so the low-energy effective theory contains three dynamical light quark flavors, $q=u,d,s$.

In addition to dipole operators such as Eq.~\eqref{eq:DipoleOperator}, LEFT has dimension-six $q^4$, $\ell^4$, $q^2 \ell^2$, and $G^3$ operators, where $q$, $\ell$, and $G$ are quark, lepton, and gluon fields. The $\ell^4$ and $q^2 \ell^2$ operators contribute to lepton-flavor-violating processes such as $\mu\to e \gamma$. The $\ell^4$ operators contribute to $\mu \to e \gamma$ via a perturbative loop graph giving $\lwc{e\gamma}{} \sim e m_\ell/(16\pi^2) \lwc{\ell^4}{}$, where $\lwc{\ell^4}{}$ is a typical $\ell^4$ operator coefficient.  In this paper, we compute the non-perturbative contribution of semileptonic $q^2 \ell^2$ operators to the purely leptonic process $\mu \to e \gamma$, and show that
these are of order $\lwc{e\gamma}{} \sim e (F_\pi^2/\lchi) \lwc{q^2\ell^2}{} $, where $\lchi \sim 4 \pi F_\pi$ is the scale of chiral symmetry breaking~\cite{Manohar:1983md}. These contributions are significant, and the limit Eq.~\eqref{eq:BRmuegamma} on the $\mu \to e \gamma$ branching ratio puts interesting limits on semileptonic tensor operators.

The constraints of lepton-flavor-violating  processes such as $\mu\to e\gamma$ and $\mu\to3e$ on the Wilson coefficients of both SMEFT and LEFT have attracted a lot of attention in connection with perturbative effects~\cite{Crivellin:2013hpa,Pruna:2014asa,Davidson:2016edt,Crivellin:2017rmk}, while the non-perturbative effects due to lepton-flavor-violating quark operators have been studied mainly in reactions involving hadrons as external states~\cite{Black:2002wh,Carpentier:2010ue,Celis:2013xja,Celis:2014asa,Crivellin:2014cta,Crivellin:2016vjc,Hazard:2017udp,Davidson:2018rqt}.
In our analysis, we consider non-perturbative contributions to $\mu \to e \gamma$, a decay that has no strongly interacting external particles.

Within the LEFT, the perturbative renormalization-group equations can be used to compute the running and mixing of operators from the weak scale down to the hadronic scale around $\sim 2\GeV$.  Below this scale, non-perturbative QCD effects are present and must be included.  At low energies, the interactions of the lightest hadrons are described by chiral perturbation theory (\chpt{})~\cite{Weinberg:1968de,Gasser:1983yg,Gasser:1984gg}. In order to express the low-energy constants of \chpt{} in terms of the LEFT parameters, a non-perturbative matching has to be performed, either using lattice QCD or phenomenological input. In this paper, we study the matching of semileptonic $q^2 \ell^2$ operators onto \chpt{} operators that contribute to processes with no hadrons.

\section{Matching the LEFT to chiral perturbation theory}\label{sec:ChPTMatching}

If we consider the leptonic sector of the LEFT, at leading order in the electromagnetic coupling, the quark dipole operators and the $q^4$ and $G^3$ operators only affect the hadronic vacuum polarization function. The most interesting non-perturbative effect on the lepton sector is given by semileptonic $q^2 \ell^2$ operators. We consider the subset of LEFT operators involving a charged-lepton bilinear and a quark bilinear~\cite{Jenkins:2017dyc}, shown in Table~\ref{tab:oplist}.

\begin{table}[t]
\begin{align*}
	\small
	\renewcommand{\arraystretch}{1.51}
	\begin{array}[t]{c|c}
	\multicolumn{2}{c}{\boldsymbol{(\overline L L)(\overline L  L)}} \\
	\hline
	\Op{eu}{LL}[V] & (\bar e_{Lp} \gamma^\mu e_{Lr})(\bar u_{Lw} \gamma_\mu u_{Lt}) \\
	\Op{ed}{LL}[V] & (\bar e_{Lp} \gamma^\mu e_{Lr})(\bar d_{Lw} \gamma_\mu d_{Lt}) \\[1.6cm]
	\multicolumn{2}{c}{\boldsymbol{(\overline R R)(\overline R  R)}} \\
	\hline
	\Op{eu}{RR}[V] & (\bar e_{Rp} \gamma^\mu e_{Rr})(\bar u_{Rw} \gamma_\mu u_{Rt}) \\
	\Op{ed}{RR}[V] & (\bar e_{Rp} \gamma^\mu e_{Rr})(\bar d_{Rw} \gamma_\mu d_{Rt}) \\
	\end{array}
	\quad %
	\begin{array}[t]{c|c}
	\multicolumn{2}{c}{\boldsymbol{(\overline L L)(\overline R  R)}} \\
	\hline
	\Op{eu}{LR}[V] & (\bar e_{Lp} \gamma^\mu e_{Lr})(\bar u_{Rw} \gamma_\mu u_{Rt}) \\
	\Op{ed}{LR}[V] & (\bar e_{Lp} \gamma^\mu e_{Lr})(\bar d_{Rw} \gamma_\mu d_{Rt}) \\
	\Op{ue}{LR}[V] & (\bar u_{Lp} \gamma^\mu u_{Lr})(\bar e_{Rw} \gamma_\mu e_{Rt}) \\
	\Op{de}{LR}[V] & (\bar d_{Lp} \gamma^\mu d_{Lr})(\bar e_{Rw} \gamma_\mu e_{Rt}) \\
	\end{array}
	\quad %
	\begin{array}[t]{c|c}
	\multicolumn{2}{c}{\boldsymbol{(\overline L R)(\overline L R) + \hc}} \\
	\hline
	\Op{eu}{RR}[S] & (\bar e_{Lp} e_{Rr})(\bar u_{Lw} u_{Rt}) \\
	\Op{ed}{RR}[S] & (\bar e_{Lp} e_{Rr})(\bar d_{Lw} d_{Rt}) \\
	\Op{eu}{RR}[T] & (\bar e_{Lp} \sigma^{\mu\nu} e_{Rr})(\bar u_{Lw} \sigma_{\mu\nu} u_{Rt}) \\
	\Op{ed}{RR}[T] & (\bar e_{Lp} \sigma^{\mu\nu} e_{Rr})(\bar d_{Lw} \sigma_{\mu\nu} d_{Rt}) \\[0.2cm]
	\multicolumn{2}{c}{\boldsymbol{(\overline L R)(\overline R L) + \hc}} \\
	\hline
	\Op{eu}{RL}[S] & (\bar e_{Lp} e_{Rr})(\bar u_{Rw} u_{Lt}) \\
	\Op{ed}{RL}[S] & (\bar e_{Lp} e_{Rr})(\bar d_{Rw} d_{Lt})
	\end{array}
\end{align*}
\caption{Semileptonic LEFT operators involving a charged-lepton bilinear and a quark bilinear.}
\label{tab:oplist}
\end{table}

Quark operators in the LEFT Lagrangian can be matched onto operators in the chiral Lagrangian, as long as the momentum transfer is well below $\lchi$. Such a matching has been extensively used in hadronic weak decays (e.g.\ see Ref.~\cite{Georgi:1985kw}). The results for scalar and vector quark bilinears can be obtained from the usual \chpt{} Lagrangian including external sources~\cite{Gasser:1983yg,Gasser:1984gg}. The extension to tensor sources has been derived in Ref.~\cite{Cata:2007ns}.

We quickly remind the reader of the construction of the chiral Lagrangian with external sources. The massless QCD Lagrangian is supplemented by quark bilinears according to
\begin{align}
	\L = \L_\mathrm{QCD}^{M=0} + \L_\mathrm{ext} =  \L_\mathrm{QCD}^{M=0} &+ \bar q_L \gamma^\mu l_\mu q_L + \bar q_R \gamma^\mu r_\mu q_R + \bar q_L S q_R +  \bar q_R S^\dagger q_L  \nn
			&+ \bar q_L \sigma^{\mu\nu} t_{\mu\nu} q_R  + \bar q_R \sigma^{\mu\nu} t_{\mu\nu}^\dagger q_L   \, ,
\end{align}
where $q = (u,d,s)^T$ is the three-component column vector of the light quarks and the external sources $r_\mu$, $l_\mu$, $S$, and $t_{\mu\nu}$ are $3\times3$ matrices in flavor space.\footnote{Compared with the notation in Ref.~\cite{Cata:2007ns}, we use $t_{\mu \nu} \to t_{\mu \nu}^\dagger$ since the LEFT basis uses $\bar q_L \sigma^{\mu \nu} q_R$ operators. The usual \chpt{} scalar source is $\chi = -2 B_0 S^\dagger$.} By promoting the sources to spurion fields transforming under chiral rotations, one obtains a QCD Lagrangian which is formally invariant under chiral symmetry, and one then constructs the most general chiral Lagrangian based on this chiral symmetry involving the spurion source fields and the matrix
\begin{align}
	U(x) &= \exp\Big( i \frac{\pi(x)}{F_0} \Big)\,, & \pi(x) &= \lambda^a \pi^a(x)\,, & \text{Tr}( \lambda^a \lambda^b ) &=2\, \delta^{ab}\,,
\end{align}
transforming as $U \to R U L^\dagger$ under chiral $SU(3) \times SU(3)$ rotations.  In the above equations, the matrix $\pi(x)$ is defined in terms of the eight Goldstone boson fields $\pi^a(x)$, $a=1, \cdots,8$, the Gell-Mann matrices $\lambda^a$, and the pion decay constant in the chiral limit, $F_0$.  Green functions of quark bilinears are computed with the chiral Lagrangian via functional derivatives with respect to the sources. Finally, the spurion sources are fixed to the physical values including the contribution of the semileptonic LEFT operators.

Since the semileptonic LEFT contributions to the sources include a factor $1/v^2$ with $v\sim246\GeV$, we are only interested in terms linear in the LEFT sources: quadratic terms are of the same size as dimension-eight contributions in the LEFT and can be neglected. This applies to weak SM contributions to the LEFT coefficients and even more so to BSM contributions suppressed by $1/\lbsm^2$. Therefore, we explicitly split the contribution to the spurions into two parts:
\begin{align}
	\label{eq:Spurions}
	S \mapsto S + \tilde S \, , \quad
	r_\mu \mapsto r_\mu + \tilde r_\mu \, , \quad
	l_\mu \mapsto l_\mu + \tilde l_\mu \, , \quad
	t_{\mu\nu} \mapsto t_{\mu\nu} + \tilde t_{\mu\nu} \, .
\end{align}
The spurions $S$, $r_\mu$, $l_\mu$, and $t_{\mu\nu}$ describe the mass matrix and the coupling to the electromagnetic field. They are fixed in the end to their physical values (in our sign convention the QED covariant derivative is given by $D_\mu = \p_\mu + i e Q A_\mu$):
\begin{align}
	S \mapsto - M_q^\dagger \, , \quad
	r_\mu \mapsto - e Q A_\mu \, , \quad
	l_\mu \mapsto - e Q A_\mu \, , \quad
	t_{\mu\nu} \mapsto 0 \, ,
\end{align}
where $M_q = \mathrm{diag}(m_u, m_d, m_s)$ is the light-quark mass matrix.

The spurions $\tilde S$, $\tilde r_\mu$, $\tilde l_\mu$, and $\tilde t_{\mu\nu}$ contain the contributions from higher-dimensional operators. We only consider terms linear in the dimension-five and dimension-six LEFT operator coefficients, e.g.\ the first entry in the $3\times3$ matrix
\begin{align}
	\tilde S = \begin{pmatrix}
		\tilde S_{uu} &  \tilde S_{ud} &  \tilde S_{us} \\
		\tilde S_{du} &  \tilde S_{dd} &  \tilde S_{ds} \\
		\tilde S_{su} &  \tilde S_{sd} &  \tilde S_{ss}
	 \end{pmatrix}
\end{align}
is given by
\begin{align}
	\tilde S_{uu} = {\lwc{eu}{RL}[S][pruu]}^* (\bar e_{Rr} e_{Lp} ) + {\lwc{eu}{RR}[S][pruu]} (\bar e_{Lp} e_{Rr}) \, .
\end{align}
Note that terms proportional to $G_F$ induced by the SM weak interaction are included in $\tilde l_\mu$.

At order $p^4$ in the chiral counting,\footnote{We will treat the LEFT sources formally as the same order in chiral counting as the usual chiral sources, $\tilde S \sim p^2$, $\tilde r_\mu \sim p$, $\tilde l_\mu \sim p$, $\tilde t_{\mu \nu} \sim p^2$, so that the equations can be compared easily with the \chpt{} literature.}
the matching is easily computed from the chiral Lagrangian by shifting sources as in Eq.~\eqref{eq:Spurions}, and amounts to the following replacement rules for the LEFT quark bilinears.\\
{\bf Scalar:}
\begin{align}
	\label{eq:Scalar}
	\bar q_L \tilde S q_R \to -2B_0 &\bigg[\frac14F_0^2 \vev{\tilde S U} +L_4  \vev{D_\mu U^\dagger D^\mu U} \vev{\tilde S U} + L_5  \vev{\tilde S UD_\mu U^\dagger D^\mu U } \nn
		& +2 L_6 \vev{U^\dagger \chi + \chi^\dagger U} \vev{\tilde SU} -2 L_7 \vev{U^\dagger \chi - \chi^\dagger U} \vev{\tilde SU} +2 L_8 \vev{ \tilde S U \chi^\dagger U} \nn
		& + H_2 \vev{\tilde S \chi}\bigg] + \O(p^6) \, .
\end{align}
{\bf Vector:}
\begin{align}
	\label{eq:Vector}
	\bar q_R \gamma^\mu \tilde r_\mu q_R &\to \frac{i}2 F_0^2 \vev{\tilde r^\mu  D_\mu U U^\dagger } + 4 i L_1 \vev{D_\nu U^\dagger D^\nu U} \vev{\tilde r^\mu D_\mu U U^\dagger  } \nn
	& + 4i L_2 \vev{D^\mu  U^\dagger D^\nu U} \vev{\tilde r_\mu D_\nu U U^\dagger }  + 2 i L_3 \vev{\left(U^\dagger \tilde r_\mu D^\mu U - D^\mu U^\dagger  \tilde r_\mu  U  \right) D_\nu U^\dagger D^\nu U }  \nn
	& + 2 i L_4 \vev{\tilde r^\mu  D_\mu U U^\dagger   } \vev{U^\dagger \chi + \chi^\dagger U}  + i L_5  \vev{ \left( U^\dagger \tilde r_\mu D^\mu U  -D^\mu U^\dagger \tilde r_\mu U  \right) (U^\dagger \chi + \chi^\dagger U)  } \nn
	& +L_9 \left[ -\vev{\tilde r_\mu F_R^{\mu \nu} D_\nu U U^\dagger}  - \vev{\tilde r_\mu U D_\nu U^\dagger F_R^{\mu \nu} } + \vev{\tilde r_\mu D_\nu U F_L^{\mu \nu} U^\dagger} + \vev{\tilde r_\mu U F_L^{\mu \nu} D_\nu U^\dagger}  \right] \nn
	& -i L_9 \vev{\tilde r^\mu D^\nu(D_\mu U D_\nu U^\dagger-D_\nu U D_\mu U^\dagger}  + 2 L_{10} \vev{\tilde r_\mu D_\nu( U F_L^{\mu \nu} U^\dagger)} \nn 
	&  + 4 H_1 \vev{\tilde r_\mu D_\nu F_R^{\mu \nu}} + \epsilon\text{ terms}+\O(p^6) \, ,
\end{align}
where we have omitted $\O(p^4)$ terms involving $\epsilon_{\alpha \beta \lambda \sigma}$ from the odd-intrinsic-parity (anomalous) sector~\cite{Wess:1971yu,Witten:1983tw,Manohar:1984uq,Kaiser:2000gs}. \\
{\bf Tensor:}
\begin{align}
	\label{eq:Tensor}
	\bar q_L \sigma^{\mu\nu} \tilde t_{\mu\nu} q_R &\to \Lambda_1 \vev{ \tilde t_{\mu \nu} (U F_L^{\mu \nu} + F_R^{\mu \nu}U) } + i \Lambda_2 \vev{ \tilde t^{\mu \nu} D_\mu U U^\dagger D_\nu U} + \O(p^6) \, .
\end{align}
Here $F_R^{\mu\nu} = \p_\mu r_\nu -\p_\nu r_\mu -i\left[r_\mu,\, r_\nu\right]$ and similarly for $F_L^{\mu\nu}$ with $r_\mu\to l_\mu$. $B_0$, $L_i$, and $H_i$ are the usual non-perturbative low-energy constants (LECs) that enter the chiral Lagrangian to order $p^2$ and $p^4$. $\Lambda_{1,2}$ are analogous parameters for tensor sources studied in Ref.~\cite{Cata:2007ns}. The matching of $\bar q_R S^\dagger q_L$ and $\bar q_R t_{\mu \nu}^\dagger q_L$ is given by the Hermitian conjugates of Eqs.~\eqref{eq:Scalar}, \eqref{eq:Tensor}, and the one of $\bar q_L l^\mu \gamma_\mu q_L$ by $r_\mu \to l_\mu$, $U \leftrightarrow U^\dagger$, $\chi \leftrightarrow \chi^\dagger$ in Eq.~\eqref{eq:Vector}. 
When computing matrix elements to order $p^4$, one has to include the contribution from one-loop diagrams with $p^2$ vertices and tree graphs with $p^4$ vertices, as usual in \chpt.

The operators on the l.h.s.~of Eqs.~\eqref{eq:Scalar}, \eqref{eq:Vector}, and \eqref{eq:Tensor} are renormalized at a scale $\mu$ at which perturbation theory is still valid. While the combination $\chi = - 2 B_0 S^\dagger$ and the LECs $L_i$ are independent of the perturbative scale $\mu$~\cite{Gasser:1983yg}, the LECs $B_0$ and $\Lambda_i$ are scale dependent~\cite{Gasser:1983yg,Cata:2007ns}. They connect the perturbative scale $\mu$ with low-energy hadron dynamics.

We can now match $q^2\ell^2$ operators in LEFT onto the chiral Lagrangian. Quarks and leptons do not interact if electromagnetic effects are turned off, so we can directly get the matching to lowest order in $\alpha$ from Eqs.~\eqref{eq:Scalar}--\eqref{eq:Tensor}. For example, the LEFT operators
\begin{align}
	\label{eq:TensorOp}
	\Op{eu}{RR}[T][prwt] &= (\bar e_{Lp} \sigma^{\mu \nu} e_{Rr}) (\bar u_{Lw} \sigma_{\mu \nu} u_{Rt}) , &  \Op{ed}{RR}[T][prwt] &= (\bar e_{Lp} \sigma^{\mu \nu} e_{Rr}) (\bar d_{Lw} \sigma_{\mu \nu} d_{Rt})  
\end{align}
are matched onto
\begin{align}
	\label{eq:TensorOpMatching}
	\Op{eq}{RR}[T][prwt]
	& \to \Lambda_1 (\bar e_{Lp} \sigma^{\mu \nu} e_{Rr})(U F_L^{\mu \nu} + F_R^{\mu \nu}U)_{tw}  \nn
	&\quad + i \Lambda_2 (\bar e_{Lp} \sigma^{\mu \nu} e_{Rr}) (D_\mu U U^\dagger D_\nu U)_{tw} + \O(p^6)
\end{align}
in \chpt, where $\Lambda_{1,2}$ are the \emph{same} as in Eq.~\eqref{eq:Tensor} up to corrections of order $\alpha$. Since we are matching onto \chpt, Eq.~\eqref{eq:TensorOpMatching} can be used for lepton flavors $p,r=e,\mu$, and quark flavors $w,t=u,d,s$. Similarly, using Eq.~\eqref{eq:Scalar}, the matching of the scalar operators
\begin{align}
	\label{eq:ScalarOp}
	\Op{eu}{RR}[S][prwt] &= (\bar e_{Lp} e_{Rr})(\bar u_{Lw} u_{Rt}), &
	\Op{ed}{RR}[S][prwt] &= (\bar e_{Lp} e_{Rr})(\bar d_{Lw} d_{Rt}),
\end{align}
is given by
\begin{align}
	\label{eq:ScalarOpMatching}
	 \Op{eq}{RR}[S][prwt] \to - 2 B_0 (\bar e_{Lp} e_{Rr}) &\bigg[ \frac14 F_0^2 U_{tw}  + L_4   \vev{D_\mu U^\dagger D^\mu U} U_{tw}  + L_5   (U D_\mu U^\dagger D^\mu U)_{tw} \nn
	 	& + 2 L_6  \vev{U^\dagger \chi + \chi^\dagger U} U_{tw}  -2 L_7  \vev{U^\dagger \chi - \chi^\dagger U} U_{tw} + 2 L_8  (U \chi^\dagger U)_{tw} \nn
		& + H_2  \chi_{tw} \bigg] + \O(p^6) \, , \quad q = u,d \, .
\end{align}
The matching of $\Op{eu}{RL}[S]$, $\Op{ed}{RL}[S]$ is given by a similar expression with the hadronic part in Eq.~\eqref{eq:ScalarOpMatching} replaced by its hermitian conjugate.

At order $\alpha$, the $q^2 \ell^2$ operators can no longer be treated as the product of quark and lepton bilinears: e.g.~the matching of the tensor operator Eq.~\eqref{eq:TensorOp} includes operators such as those in Eq.~\eqref{eq:ScalarOpMatching} where the leptonic part is a Lorentz scalar, since both operators transform as $(\bf{\overline{3}}, \bf{3})$ under chiral $SU(3) \times SU(3)$.

We want to study the effect of $q^2\ell^2$ operators on $\mu \to e \gamma$. The matrix element $\braket{\gamma(p,\epsilon) | S | 0}$ of a scalar operator to a physical photon vanishes due to Lorentz and gauge invariance. The matrix element $\braket{\gamma(p,\epsilon) | V^\mu | 0}$ of a vector operator to a physical photon has the form
\begin{align}
	\label{eq:VectorMatrixEl}
	\braket{\gamma(p,\epsilon) | V^\mu | 0} &= A(p^2) \left[ p^\mu (p \cdot \epsilon) - p^2 \epsilon^\mu \right] 
\end{align}
by gauge invariance, and so vanishes for an on-shell photon. Eq.~\eqref{eq:VectorMatrixEl} has the structure of a penguin amplitude, and does contribute if the virtual photon is attached to external fermion lines, as in $\mu \to 3e$. The lowest-order version of this argument is a consequence of Eq.~\eqref{eq:Scalar} with $U=1$ having no $F_{\mu\nu}$ term,  and Eq.~\eqref{eq:Vector} with $U=1$ being proportional to $\partial^\nu F_{\mu \nu}$. The absence of a one-photon matrix element for scalar and vector operators holds to all orders in the chiral expansion, but to lowest order in $\alpha$, since photon exchange between the quark and lepton bilinears can change the Lorentz structure of the bilinears.

\begin{figure}
\begin{center}
\includegraphics[width=2cm,angle=90]{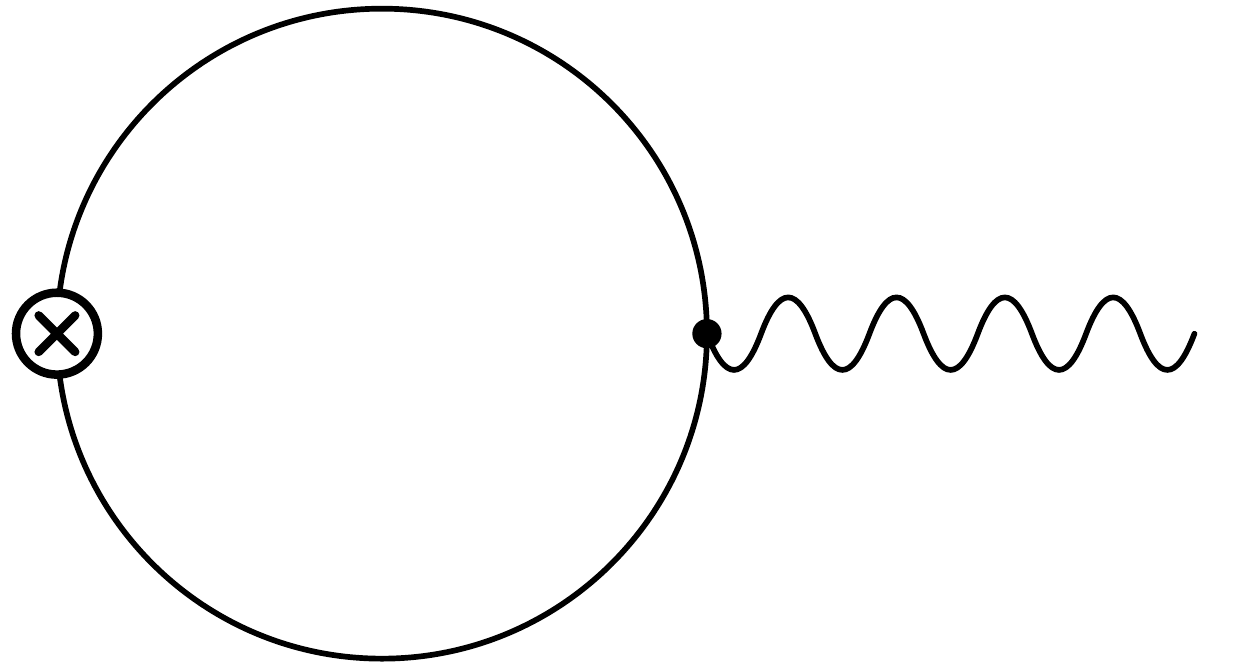}
\end{center}
\caption{\label{fig:1} The matrix element $\braket{\gamma | \bar q \sigma^{\mu \nu} q | 0}$.}
\end{figure}

Ignoring QED corrections, which can mix Lorentz structure,  the only LEFT operators that contribute to $\mu \to e \gamma$ are tensor operators, which can have a non-zero hadronic matrix element to an on-shell photon:
\begin{align}
(\bar e_{Lp} \sigma^{\mu \nu} e_{Rr}) (\bar q_{L} \sigma_{\mu \nu} q_{R})   
& \to -2 Q_q e\, \Lambda_1 \, (\bar e_{Lp} \sigma^{\mu \nu} e_{Rr})F^{\mu \nu} + \O(p^6) \, , & q&=u,d,s
\label{2.7}
\end{align}
where $F^{\mu \nu}$ is the photon field-strength tensor.
To study the impact of Eq.~\eqref{2.7}, we need the value of the non-perturbative parameter $\Lambda_1$. Naive dimensional analysis (NDA)~\cite{Manohar:1983md,Gavela:2016bzc} gives the estimate
\begin{align}
\Lambda_1 \sim c_T \frac{F_\pi^2}{\lchi} = c_T \frac{\lchi}{16 \pi^2}
\label{2.8}
\end{align}
with $c_T$ of order one. Here $F_\pi = 92.3(1)\MeV$ is the physical pion decay constant~\cite{Tanabashi:2018oca}, and $F_0 = F_\pi$ to lowest order in the chiral expansion. Eq.~\eqref{2.8} is the size one would expect from the graph in Fig.~\ref{fig:1} where the loop integral is estimated using the scale $\lchi$.
The model estimate of Ref.~\cite{Mateu:2007tr} gives $c_T \approx -3.2$. If we use the lattice input from Ref.~\cite{Baum:2011rm} for the pion tensor charge and relate $\Lambda_1$ and $\Lambda_2$ by vector-meson saturation~\cite{Ecker:1988te}, we obtain $c_T \approx -1.0(2)$, where the error is due to uncertainties in the vector couplings $F_V$ and $G_V$ and does not include model uncertainties. Similar estimates can be obtained from~\cite{Cata:2008zc,Miranda:2018cpf}. From Eq.~\eqref{2.7} we obtain the low-energy matching condition
\begin{align}
\delta \lwc{e\gamma}{}[][e\mu] &=e\,c_T \frac{F_\pi^2}{\lchi}\left[ \frac23 \lwc{ed}{RR}[T][e\mu dd] +\frac23 \lwc{ed}{RR}[T][e\mu ss] -  \frac43 \lwc{eu}{RR}[T][e\mu uu]  \right]
\label{2.9}
\end{align}
for the additional non-perturbative contribution to the dipole operator,
where the LEFT coefficients $\wc{eq}{T}{RR}$ are evaluated at a low scale $\mu \sim 2\GeV$, and $c_T(\mu)$ is computed from tensor operators renormalized at the same scale. Note that in~\eqref{2.9}, the $\mu$ dependence due to gluon corrections cancels in the product of the LEC $c_T$ and the tensor operator coefficients.

Higher-order corrections in the chiral expansion to~\eqref{2.9} could be calculated~\cite{Cata:2007ns} but would involve even more low-energy constants. For off-shell photons, the momentum dependence of tensor-current matrix elements can be determined via unitarity and dispersion relations~\cite{Hoferichter:2018zwu}.

\section{Constraints on electromagnetic dipole interactions}\label{sec:Constraints}

\subsection{Lepton-flavor-violating processes}
\label{sec:LFVConstraints}

For notational simplicity, we define
\begin{align}
L^T_{e \mu} &=  \frac23 \lwc{ed}{RR}[T][e\mu dd] +\frac23 \lwc{ed}{RR}[T][e\mu ss] -  \frac43 \lwc{eu}{RR}[T][e\mu uu] \, ,  &
L^T_{\mu e} &=  \frac23 \lwc{ed}{RR}[T][\mu e dd] +\frac23 \lwc{ed}{RR}[T][\mu e ss] -  \frac43 \lwc{eu}{RR}[T][\mu e uu] \,  ,
\label{2.11}
\end{align}
which contribute to $\mu \to e \gamma$ using Eq.~\eqref{2.9}.
Eq.~\eqref{eq:LimitWilsonCoeff} gives the bound
\begin{align}
\abs{c_T }\left( \abs{L^T_{e \mu}}^2 + \abs{L^T_{\mu e}}^2 \right)^{1/2} \lesssim 1.65 \times 10^{-5} \TeV^{-2} \, ,
\label{2.10}
\end{align}
using Eq.~\eqref{2.9} as the only contribution to the dipole operator, which translates to $\lbsm > \abs{c_T}^{1/2} 250\TeV \approx 450\TeV$, using $\abs{c_T}=3.2$.

In Table~\ref{tab:2GeV}, we collect the constraints on the dipole and tensor operator coefficients at the hadronic scale $\mu=2\GeV$. 
Apart from the above constraints from $\mu\to e\gamma$, the same tensor operators  contribute to $\mu\to e$ conversion in nuclei. Due to their Lorentz structure these operators  induce spin-dependent $\mu\to e$ amplitudes, and do not directly contribute to the  $A^2$-enhanced spin-independent rate. However, these interactions do indirectly induce a spin-independent amplitude through their contribution to the dipole operators in Eq.~\eqref{2.9}.\footnote{It should be noted that the spin-dependent contributions in principle contribute at a similar level as the spin-independent ones induced through the dipole operators. Furthermore, there are additional non-perturbative contributions as discussed in footnote~\footref{fnt:AdditionalContributions}. Here we simply take the contributions arising from the dipole operators as an estimate.}
As a result, we can compare the limits on the tensor operators from $\mu\to e\gamma$ and $\mu\to e$ conversion, the resulting bounds from the latter are shown in the last two rows of Table~\ref{tab:2GeV}.
Here we used the expressions of Refs.~\cite{Cirigliano:2009bz,Crivellin:2017rmk}, which are collected in App.~\ref{App:conv}, together with the current and projected experimental constraints~\cite{Bertl:2006up,Cui:2009zz,Kutschke:2011ux,Kuno:2013mha}
\begin{align}
{\rm Br}({\rm Au}) \leq 7\times 10^{-13},\qquad   {\rm Br}({\rm Al}) \lesssim 10^{-16}\qquad ({\rm projected})\,,
\label{eq:ConvExpt}
\end{align}
and employed the same hadronic and nuclear input as Ref.~\cite{Crivellin:2017rmk}.
The constraints in Table~\ref{tab:2GeV} are derived by turning on a single operator at a time at the scale $\mu=2\GeV$. In this scenario, the current constraints from $\mu\to e\gamma$ are stronger than the constraints from $\mu\to e$ conversion, while the projected bounds are comparable. Note, however, that in a realistic BSM scenario, there is no reason to believe that only a single operator is present at the hadronic scale and in general the constraints have to be applied to the linear combination $\lwc{e\gamma}{} + \delta\lwc{e\gamma}{}$.

In a second step, we consider the constraints on the operator coefficients at the scale $\mu=M_W$ by taking into account the renormalization-group equations (RGEs) discussed in App.~\ref{app:rge}. Detailed analyses of these RGEs have been presented in Refs.~\cite{Crivellin:2016ebg,Davidson:2018rqt,Carpentier:2010ue,Davidson:2016edt,Crivellin:2017rmk,Davidson:2017nrp,Cirigliano:2017azj}. We evaluate the bounds from $\mu\to e\gamma$ and $\mu\to e$ conversion by turning on single operators at a time at the scale $\mu=M_W$.
These constraints almost directly translate to limits on operators in the SMEFT, using the tree-level matching~\cite{Jenkins:2017jig},
\begin{align}\label{eq:3.4}
\wc{eu}{S}{RR} = - C_{lequ}^{(1)} \, ,\qquad 
\wc{eu}{T}{RR} = - C_{lequ}^{(3)} \, .
\end{align}
In the case of the tensor operators, the running and mixing from $\mu=M_W$ down to $\mu=2\GeV$ and the non-perturbative matching lead to two effects at the hadronic scale: on the one hand, the tensor operators contribute to the effective dipole operators through the non-perturbative matching effects discussed in Sect.~\ref{sec:ChPTMatching}, leading to a contribution proportional to $c_T$. 
On the other hand, the RGE evolution from $\mu=M_W$ down to $\mu=2\GeV$ induces perturbative mixing contributions to the coefficients of the dipole operators which are independent of $c_T$. Therefore, the bounds on $\mu\to e\gamma$ constrain the linear combinations of terms proportional of $c_T$ and independent of $c_T$ as shown in the upper half of Table~\ref{tab:Mw}.  Assuming $c_T=\O(1)$, the non-perturbative contribution  dominates for the couplings to up and down quarks, while it is expected to be of the same order of magnitude as the perturbative term for the strange-quark couplings.

In the case of $\mu\to e$ conversion, the RGEs have more important consequences: the tensor operator coefficients also mix into the coefficients of semileptonic scalar operators $\lwc{ed,eu}{RR}[S]$~\cite{Crivellin:2017rmk,Jenkins:2017dyc}. As discussed in Ref.~\cite{Cirigliano:2017azj}, these mixing effects from $\mu=M_W$ down to $\mu=2\GeV$ turn out to be large and give a significant contribution to the spin-independent amplitude.\footnote{\label{fnt:AdditionalContributions}
In principle there are additional contributions to the spin-independent amplitude that also appear at $\O(e^2/(4\pi)^2)$. Firstly,  a spin-independent $(\bar NN)(\bar \mu e)$ interaction can be induced in the chiral Lagrangian by  the combination of the tensor operator with two electromagnetic currents. In addition, there are chiral loops involving photon exchange around the leading-order spin-dependent $(\bar NN)(\bar \mu e)$ interaction that is generated by $ \wc{ed,eu}{T}{RR}$. As these contributions involve unknown LECs we neglect them here and use the RGE-induced contribution as an estimate of the full amplitude.} 
As in the case of $\mu\to e\gamma$, the bounds on $\mu\to e$ conversion constrain a combination of the couplings at $\mu=M_W$ that are independent of $c_T$ and a non-perturbative contribution proportional to $c_T$. Due to the large mixing into scalar operators, the non-perturbative pieces are negligible for the couplings to the up and down quarks, while they contribute at the $\O(10\%)$ level for the strange-quark couplings, as shown in the lower half of Table~\ref{tab:Mw}.

Note that the current constraints from $\mu\to e $ conversion are of the same order of magnitude  as the limits from $\mu\to e\gamma$, if we use the model estimate $c_T\approx -3$. In this case, the $\mu\to e$ conversion limits  are somewhat stronger for the couplings involving the up and down quarks, while the strange-quark couplings are more stringently constrained by the $\mu\to e\gamma$ limit. It should be mentioned that these comparisons are currently rather uncertain due to the fact that only NDA and model estimates exist for $c_T$.
In addition,  the two observables depend on different linear combinations of the tensor interactions. In particular, the combination of couplings that enters $\mu\to e\gamma$  only depends on the quark charges, while the spin-independent $\mu\to e$ conversion rate also involves the matrix elements $\langle N| \bar q q|N\rangle$. This implies that the two observables are complementary probes, as one would generally expect BSM physics to generate multiple operators in the LEFT.

\begin{table}[t]
\begin{align*}
	\small
	\renewcommand{\arraystretch}{1.51}\arraycolsep=6pt
	\begin{array}[t]{ll|ccccccc}
	\toprule
	\multicolumn{2}{l|}{\mu=2\GeV} & \lwc{e\gamma}{}[][e\mu] & c_T L_{\substack{eu\\ e\mu uu}}^{T,RR} & c_T L_{\substack{ed\\ e\mu dd}}^{T,RR} & c_T L_{\substack{ed\\ e\mu ss}}^{T,RR}\\
	\midrule
	\text{(c)} & {\rm Br}(\mu\to e \gamma) \leq 4.2\times 10^{-13}	& 3.7\times 10^{-11} &1.2 \times 10^{-5} &  2.5  \times 10^{-5} & 2.5 \times 10^{-5}\\
	\text{(e)} & {\rm Br}(\mu\to e \gamma) \lesssim 4\times 10^{-14} &1.1\times 10^{-11} &3.8 \times 10^{-6} &  7.6  \times 10^{-6} & 7.6 \times 10^{-6}\\
	\midrule
	\text{(c)} & {\rm Br}({\rm Au})\leq 7\times 10^{-13}	& 7.6\times 10^{-10} &2.6 \times 10^{-4} &  5.1  \times 10^{-4} & 5.1 \times 10^{-4}\\
	\text{(e)} & {\rm Br}({\rm Al})\lesssim 1\times 10^{-16}	&1.1\times 10^{-11} &3.7 \times 10^{-6} &  7.4  \times 10^{-6} & 7.4 \times 10^{-6} \\
	\bottomrule
	\end{array}
		\end{align*}
		\caption{Current (c) and expected (e) limits from $\mu \to e\gamma$ and $\mu \to e$ conversion experiments on the coefficients of the dipole and tensor operators at the scale $\mu=2\GeV$. The limits are given in units of $\TeV^{-2}$ for the tensor operators, while those for $ \lwc{e\gamma}{}[][e\mu]$ are in units of $\TeV^{-1}$. They constrain the absolute values of both the listed coefficients as well as the coefficients $\lwc{e\gamma}{}[][\mu e]$ and $c_T \lwc{eq}{RR}[T][\mu e qq]$.}
	\label{tab:2GeV}
\end{table}

\begin{table}[t]
\begin{align*}
	\small
	\renewcommand{\arraystretch}{1.51}\arraycolsep=6pt
	\scalebox{0.83}{
	\begin{array}[t]{ll|ccccccc}
	\toprule
	\multicolumn{2}{l|}{\mu=M_W} & \lwc{e\gamma}{}[][e\mu]&(c_T-0.07) L_{\substack{eu\\ e\mu uu}}^{T,RR} &(c_T-0.15) L_{\substack{ed\\ e\mu dd}}^{T,RR} & (c_T-3.1) L_{\substack{ed\\ e\mu ss}}^{T,RR}\\
	\midrule
	\text{(c)} & {\rm Br}(\mu\to e \gamma) \leq 4.2\times 10^{-13}&3.8\times 10^{-11} &1.4 \times 10^{-5} &  2.8  \times 10^{-5} & 2.8 \times 10^{-5}\\
	\text{(e)} & {\rm Br}(\mu\to e \gamma) \lesssim 4\times 10^{-14}&1.2\times 10^{-11} &4.4 \times 10^{-6} &  8.8  \times 10^{-6} & 8.8 \times 10^{-6}\\
	\midrule
	&& \lwc{e\gamma}{}[][e\mu]&(1+5\times 10^{-3}c_T) L_{\substack{eu\\ e\mu uu}}^{T,RR} & (1+5\times 10^{-3}c_T) L_{\substack{ed\\ e\mu dd}}^{T,RR} &(1+0.1c_T)  L_{\substack{ed\\ e\mu ss}}^{T,RR}\\
	\midrule
	\text{(c)} & {\rm Br}({\rm Au})\leq 7\times 10^{-13}&7.9\times 10^{-10} &1.6 \times 10^{-6} &  3.0  \times 10^{-6} & 7.0 \times 10^{-5}\\
	\text{(e)} & {\rm Br}({\rm Al})\lesssim 1\times 10^{-16}&1.1\times 10^{-11} &2.0 \times 10^{-8} &  4.0  \times 10^{-8} & 8.9\times 10^{-7} \\
	\bottomrule
	\end{array}
	}
		\end{align*}
		\caption{Current (c) and expected (e) limits from $\mu \to e\gamma$ and $\mu \to e$ conversion experiments on the coefficients of the dipole and tensor operators at the scale $\mu=M_W$. The first and fourth rows indicate the combinations of Wilson coefficients and the LEC $c_T$ that are constrained by each observable. The limits are given in units of TeV$^{-2}$ for the tensor operators, while those for $ \lwc{e\gamma}{}[][e\mu]$ are in units of TeV$^{-1}$. The linear combinations $ (1+5\times 10^{-3}c_T)$ etc.\ differ by about 10\% between Au and Al. }
	\label{tab:Mw}
\end{table}
	
\subsection{Constraints on flavor-diagonal dipole moments}
\label{sec:EDMs}

Analogously to the $\mu\to e\gamma$  operators in  Eq.~\eqref{2.9}, flavor-diagonal dipole operators can be induced by $\lwc{eu}{T,RR}[][lluu]$, $\lwc{ed}{T,RR}[][lldd]$, and $\lwc{ed}{T,RR}[][llss]$.
These dipole couplings subsequently contribute to the anomalous magnetic moments of the leptons, $\Delta a_l = a_l -a_l^{\rm SM}$, through their real parts, while the imaginary parts induce electric dipole moments (EDMs), $d_l$,
\begin{align}
	\Delta a_l = -4\,\frac{m_l}{e}{\rm Re}\, \lwc{e\gamma}{}[][ll], \qquad d_l = -2 \,{\rm Im}\, \lwc{e\gamma}{}[][ll]\,.
\end{align}
With the most recent measurement of the fine-structure constant~\cite{Parker2018} a tension of $2.4\sigma$ between SM prediction~\cite{Aoyama:2017uqe} and measurement~\cite{Hanneke:2008tm} of the anomalous magnetic moment of the electron is observed:
\begin{align}
	\Delta a_e = -0.88(36) \times 10^{-12} \, .
\end{align}
In the case of the muon, there is a long-standing discrepancy of $3$--$4\sigma$ between the experimental value~\cite{Bennett:2006fi} and the SM prediction, which consists of QED~\cite{Laporta:1996mq,Laporta:2017okg,Aoyama:2017uqe}, electroweak~\cite{Jackiw:1972jz,Altarelli:1972nc,Bars:1972pe,Fujikawa:1972fe,Peris:1995bb,Czarnecki:1995wq,Gribouk:2005ee,Gnendiger:2013pva}, and hadronic~\cite{Prades:2009tw,Kurz:2014wya,Colangelo:2014qya,Nyffeler:2017ohp,Davier:2017zfy,Jegerlehner:2017lbd,Keshavarzi:2018mgv} contributions:\footnote{A broad effort is being made to corroborate and reduce the hadronic uncertainties in the vacuum-polarization~\cite{Ananthanarayan:2016mns,Chakraborty:2016mwy,DellaMorte:2017dyu,Borsanyi:2017zdw,Blum:2018mom,Colangelo:2018mtw} and light-by-light-scattering contributions~\cite{Green:2015sra,Gerardin:2016cqj,Blum:2016lnc,Colangelo:2017qdm,Colangelo:2017fiz,Blum:2017cer,Hoferichter:2018dmo,Hoferichter:2018kwz} using lattice QCD and dispersion relations.}
\begin{align}
	\Delta a_\mu = 2.64(0.63)(0.46)\times 10^{-9} \, .
\end{align}
Assuming Gaussian distributions, we translate these values to bounds 
\begin{align}
	| \Delta a_e | &\le 1.34 \times 10^{-12} \, , \quad \mathrm{CL}=90\%\, , \nn
	| \Delta a_\mu | &\le 3.64 \times 10^{-9} \, , \quad \mathrm{CL}=90\%\, ,
\end{align}
which imply limits on the tensor operators due to the induced dipole interactions. Assuming $\lwc{i}{T,RR}[][]=1/\lbsm^2$ and considering only the contribution of a single operator at $\mu=M_W$, we obtain $\lbsm\gtrsim 5$--$7\TeV$ for the tensor couplings to the electron, while for those to the muon we find $\lbsm\gtrsim1$--$2\TeV$. However, these constraints are not competitive with current LHC limits, which, in the case of $\lwc{eu}{T,RR}[][eeuu]$, find $\lbsm\gtrsim 12\TeV$~\cite{Alioli:2018ljm,Gupta:2018qil}. 

The CP-odd parts of the dipole moments can be constrained by EDM measurements, which, in the case of the muon~\cite{Bennett:2008dy} and tau~\cite{Inami:2002ah} do not lead to significant limits. The electron EDM is more relevant and is currently most stringently constrained by the recently improved measurements on the paramagnetic ThO molecule~\cite{Baron:2013eja,Andreev:2018ayy}. This system receives contributions from the electron EDM as well as from scalar interactions between electrons and nucleons~\cite{doi:10.1063/1.4968597,Skripnikov,Skripnikov:2013}
\begin{align}
\omega_{\rm ThO} =&\left[ (120.6\pm 4.9)\left(\frac{d_e}{10^{-27} e\, {\rm cm}}\right) +(181.6\pm 7.3)\times 10^7\left(\frac{Z}{A} C_S^{(p)}+\frac{A-Z}{A}C_S^{(n)}\right)\right] {\rm mrad/s}\,,\nn
C_S^{(N)} =& -v^2 {\rm Im}\left[ f_{S,N}^{(u)}\frac{m_N}{m_u} \left(\lwc{eu}{S,RR}[][eeuu]+\lwc{eu}{S,RL}[][eeuu]\right)+\sum_{q=d,s}f_{S,N}^{(q)}\frac{m_N}{m_q} \left(\lwc{ed}{S,RR}[][eeqq]+\lwc{ed}{S,RL}[][eeqq]\right)\right]\,,
\end{align}
where $f_{S,N}^{(q)}$ are the scalar couplings of the nucleons for which we use the results of Refs.~\cite{Hoferichter:2015dsa,Junnarkar:2013ac}. The tensor operators thus contribute to $\omega_{\rm ThO}$  non-perturbatively via $d_e$, while RGE contributions arise through both  $d_e$ and $C_S$ (see App.~\ref{app:rge} for details).\footnote{
Analogous to the case of $\mu \to e $ conversion, there are additional contributions that appear at the same order, $\O(e^2/(4\pi)^2)$. These  consist of a spin-independent $(\bar NN)(\bar e e)$ interaction that is induced by  the combination of the tensor operator with two electromagnetic currents, as well as chiral loops involving photon exchange around the leading-order spin-dependent $(\bar NN)(\bar e e)$ interaction. We again neglect these contributions and use the RGE-induced contribution as an estimate.
} 
In terms of the couplings at $\mu=M_W$ we have
\begin{align}
\frac{\omega_{\rm ThO}}{{\rm mrad/s}} ={\rm Im}\left[
(16+4.0\,c_T)\lwc{eu}{T,RR}[][eeuu]-(8.0+2.0\,c_T)\lwc{ed}{T,RR}[][eedd]+(5.7-2.0\, c_T)\lwc{ed}{T,RR}[][eess]
\right]\times 10^8 \TeV^2 \,,
\end{align}
showing that the non-perturbative contributions are of similar size as those induced through the RGEs. Using the experimental upper bound, $\omega_{\rm ThO}\leq 1.3$~mrad/s \cite{Andreev:2018ayy}, we obtain $\Lambda_{\rm BSM} \gtrsim 5 \times 10^3\TeV$ and $\Lambda_{\rm BSM} \gtrsim 3 \times 10^3\TeV$ for the couplings to up and down quarks, respectively, while for the interaction with strange quarks we have $\Lambda_{\rm BSM} \gtrsim 10^4\TeV$. 

The EDM of Mercury receives direct contributions from  tensor interactions as it is a diamagnetic system. Following the analysis of~\cite{Dekens:2018bci} for the atomic~\cite{Fleig:2018bsf} and hadronic~\cite{Gupta:2018lvp} input and using the current experimental constraint~\cite{Graner:2016ses,Griffith:2009zz} leads to somewhat stronger constraints on the up and down quark couplings. In particular, for the coupling to the up (down) quark, the mercury EDM leads to $\Lambda_{\rm BSM}\gtrsim 6\,(12)\times 10^3\TeV$, while it gives rise to a weaker limit on the coupling to the strange quark, $\Lambda_{\rm BSM}\gtrsim 2 \times 10^3\TeV$. This implies that although the tensor interactions contribute to the mercury EDM directly, the contributions to paramagnetic molecular systems can be significant, in particular in the case of  strange-quark couplings.

\subsection{Comment on dark-matter interactions}\label{sec:DM}

A very similar situation to the semileptonic tensor operators arises in the context of dark-matter interactions. For dark matter consisting of weakly interacting massive particles that couple to SM particles only through a heavy mediator, an effective field theory can be used to describe the dark matter interaction with SM particles~\cite{Goodman:2010ku,Hoferichter:2015ipa,Bishara:2017pfq,Brod:2017bsw}. Under the assumption that the dark matter particles carry a conserved quantum number, the dark matter fields always appear in bilinears and give rise to operators that are similar to those in the LEFT.
For example, at dimension five the dipole operators
\begin{align}
	\label{eq:DMDipole}
\mathcal{L}^{(5)}_\mathrm{DM}=	C^{(5)} \bar\chi_L \sigma^{\mu\nu} \chi_R F_{\mu\nu} + \hc
\end{align}
arise, where $\chi$ denotes the dark matter fermion. At dimension six, the EFT contains four-fermion operators $(\bar \chi \Gamma \chi) (\bar q \Gamma q)$, e.g.\ the tensor operators
\begin{align}
	\label{eq:DMTensor}
\mathcal{L}^{(6)}_\mathrm{DM}=	C^{(6)}_q (\bar\chi_L \sigma^{\mu\nu} \chi_R) (\bar q_L \sigma_{\mu\nu} q_R) + \hc
\end{align}
For direct-detection experiments, the interaction of dark matter with nucleons has to be studied. The constraints of chiral symmetry have been analyzed in~\cite{Cirigliano:2012pq,Hoferichter:2015ipa} for the case of vector, axial-vector, scalar, and pseudo-scalar interactions. The tensor operators~\eqref{eq:DMTensor} lead to spin-dependent interactions that are not enhanced by $A^2$. The leading contribution to the nucleon tensor form factors~\cite{Weinberg:1958ut,Adler:1975he} is given by the tensor charges and requires lattice QCD input~\cite{Gupta:2018lvp}. Through perturbative RGE mixing, the tensor operators induce the electromagnetic dipole operators~\eqref{eq:DMDipole}~\cite{Haisch:2013uaa}, which leads to a spin-independent contribution to the cross section~\cite{Banks:2010eh,Barger:2010gv}. In addition to these perturbative RGE effects, again a non-perturbative contribution is generated in analogy to the semileptonic dipole operators discussed in Sect.~\ref{sec:ChPTMatching}.
Given the similarities to the case of $\mu\to e\gamma$, the linear combination of the contributions due to the RGEs and those due to $c_T$ is the same as those in the first row of Table \ref{tab:Mw} (where $\lwc{e\gamma}{}[][e\mu]$ and $\lwc{ed,eu}{T,RR}[][e\mu qq]$ are to be identified with $C^{(5)}$ and $C^{(6)}_q$, respectively). As a result, for $c_T=\O(1)$, the contributions of the couplings to up and down quarks will be dominated by the non-perturbative piece, while it is expected to be of similar size as the RGE term for the strange-quark couplings.

Apart from the contribution proportional to $c_T$, another non-perturbative contribution is given by the coupling of the tensor current to pion exchange between two nucleons or to pion loops at the one-nucleon level.\footnote{We thank M.~Hoferichter for pointing this out.} The coupling of the tensor current to the pion is parametrized by the LEC $\Lambda_2$ in~\eqref{eq:Tensor}, which can be estimated using the lattice result~\cite{Baum:2011rm} for the pion tensor charge, $\Lambda_2 = 6.0(3)\MeV$.

\section{Conclusion}\label{sec:Conclusion}
In this paper, we have discussed the non-perturbative contributions of semileptonic LEFT operators to the purely leptonic process $\mu\to e\gamma$. Although scalar and vector operators do not induce this process (without additional QED corrections), $q^2\ell^2$  tensor interactions do generate this decay through a non-perturbative contribution to the leptonic dipole operators, $\lwc{e\gamma}{}\sim e(F_\pi^2/\lchi)\lwc{q^2\ell^2}{}$. To quantify this contribution we have matched the semileptonic tensor operators to \chpt{}, and estimated the required low-energy constant using NDA and the model estimate of Ref.~\cite{Mateu:2007tr}. These estimates show that the non-perturbative contribution is expected to dominate over the perturbative RGE effects for the couplings to up and down quarks, while it is of the same order of magnitude for the coupling to the strange quark. The constraints on the scale of BSM physics that can be set using the experimental limit on the $\mu\to e\gamma$ branching ratio are of the order of $\lbsm \gtrsim 450\TeV$, as discussed in Sect.~\ref{sec:Constraints}. These limits are of similar size as those obtained from $\mu\to e$ conversion in nuclei for the couplings to up and down quarks, while they currently provide the most stringent constraints on the couplings to strange quarks. 

\section*{Acknowledgements}
We thank Zvi Bern and the Mani L.~Bhaumik Institute at UCLA for a stimulating email announcement that started this investigation, and V.~Cirigliano, A.~Crivellin, M.~Hoferichter, and S.~Davidson for useful discussions and comments on the manuscript.
This work was supported in part by DOE Grant No.\ DE-SC0009919.
P.~S.\ is supported by a grant of the Swiss National Science Foundation (Project No.\ P300P2\!\_\!167751).

\appendix

\section{\boldmath $\mu\to e$ conversion in nuclei}\label{App:conv}
Here we discuss in more detail the contributions of the operators in Table~\ref{tab:oplist}  to $\mu\to e$ conversion in nuclei. The rate for this process can be written as~\cite[Eq.~(3.5)]{Crivellin:2017rmk}
\begin{align}
	\label{eq:conv}
	\Gamma(\mu A\to e A) &= \frac{m_\mu^3}{4}\bigg| D_A L_{\substack{e\gamma \\\mu e}}^*+4m_\mu \sum_{\substack{q=u,d,s\\ N=n,p}}\left(C^{SL}_q\frac{m_N}{m_q} f_{S,N}^{(q)}S^{(N)}_A+C_q^{VR}f_{V,N}^{(q)}V_A^{(N)}\right)\bigg|^2\nonumber\\
	&\quad+ \frac{m_\mu^3}{4}\bigg| D_A L_{\substack{e\gamma \\ e\mu}}+4m_\mu \sum_{\substack{q=u,d,s\\ N=n,p}}\left(C^{SR}_q\frac{m_N}{m_q} f_{S,N}^{(q)}S^{(N)}_A+C_q^{VL}f_{V,N}^{(q)}V_A^{(N)}\right)\bigg|^2\,.
\end{align}
Here $f_{S,N}^{(q)}$ and $f_{V,N}^{(q)}$ are hadronic matrix elements, related to the scalar- and vector-charges of the nucleons, respectively. $D_A$, $S_A^{(N)}$, and $V_A^{(N)}$ involve overlap integrals between the nuclear and lepton wavefunctions and are taken from ~\cite[Eq.~(3.10)]{Crivellin:2017rmk}, while $C_q^{SL,SR}$ and $C^{VL,VR}_q$ are the following combinations of couplings:
\begin{align}
	C_u^{SR} &= \lwc{eu}{RR}[S][e\mu uu] + \lwc{eu}{RL}[S][e\mu uu] \,,\quad & C_{d}^{SR} &= \lwc{ed}{RR}[S][e\mu dd]  + \lwc{ed}{RL}[S][e\mu dd]  \,,\quad & C_{s}^{SR} &= \lwc{ed}{RR}[S][e\mu ss] + \lwc{ed}{RL}[S][e\mu ss] \,,\nn
	C^{VR}_u &= \lwc{eu}{RR}[V][e\mu uu] + \lwc{ue}{LR}[V][uu e\mu] \,,\quad & C^{VR}_d &= \lwc{ed}{RR}[V][e\mu dd] + \lwc{de}{LR}[V][dd e\mu] \,,\quad & C^{VR}_s &= \lwc{ed}{RR}[V][e\mu ss] + \lwc{de}{LR}[V][ss e\mu] \,,\nn
	C^{VL}_u &= \lwc{eu}{LL}[V][e\mu uu] + \lwc{eu}{LR}[V][e\mu uu] \,,\quad & C^{VL}_d &= \lwc{ed}{LL}[V][e\mu dd] + \lwc{ed}{LR}[V][e\mu dd] \,,\quad & C^{VL}_s &= \lwc{ed}{LL}[V][e\mu ss] + \lwc{ed}{LR}[V][e\mu ss] \,.
\label{eq:ConvCouplings}
\end{align}
The couplings $C_q^{SL}$ can be obtained from $C^{SR}_q$ by replacing the labels $e\mu qq \to \mu e qq$ and taking the complex conjugate. It is conventional to define a branching ratio through ${\rm Br}(A) = \Gamma(\mu A\to eA)/\Gamma^{\rm capt}(A)$, where $\Gamma^{\rm capt}$ is the muon capture rate.
We follow Ref.~\cite{Crivellin:2017rmk} for the hadronic~\cite{Hoferichter:2015dsa,Junnarkar:2013ac} and nuclear input~\cite{Kitano:2002mt}. 

In order to obtain the non-perturbative tensor operator contribution to $\mu\to e$ conversion, one has to substitute $L_{e\gamma}\mapsto L_{e\gamma} + \delta L_{e\gamma}$ with $\delta L_{e\gamma}$ given in~\eqref{2.9}.

\section{Anomalous dimensions}\label{app:rge}

The RGEs for the dipole, scalar and tensor operators relevant for $\mu \to e \gamma$ from Ref.~\cite{Jenkins:2017dyc,Jenkins:2017jig} are summarized below. We treat $L_{e \gamma} \sim e m_\mu / \lbsm^2$ so that terms quadratic in $L_{e \gamma}$ have been dropped. These equations agree with earlier results in Ref.~\cite{Crivellin:2017rmk}. There are some differences with the results in Refs.~\cite{Davidson:2016edt,Crivellin:2016ebg,Davidson:2017nrp}.
\begin{align}
	 \dlwc{e\gamma}{}[][rs] &=  \Bigl[ \left(10 -b_{0,e}\right) e^2 \Bigr] \lwc{e\gamma}{}[][rs]  
		+ 8  e   [M_d]_{wv} \lwc{e d}{RR}[T][rsvw] - 16 e  [M_u]_{wv} \lwc{e u}{RR}[T][r s v w] - 2 e  [M_e]_{wv} \lwc{e e}{RR}[S][rwvs]    \, , \nn
\nnn
	\dlwc{eu}{RR}[S][prst] &= - \left[ \frac{26}{3} e^2  + 6 g^2 C_F \right] \lwc{eu}{RR}[S][prst] + 64 e^2  \lwc{eu}{RR}[T][prst]  \, , \nn
\nnn
	\dlwc{eu}{RR}[T][prst] &= \frac43 e^2 \lwc{eu}{RR}[S][prst] + \left[ \frac{26}{9} e^2+ 2 g^2 C_F \right] \lwc{eu}{RR}[T][prst]   \, , \nn
\nnn
	\dlwc{eu}{RL}[S][prst] &= - \left[ \frac{26}{3} e^2 + 6 g^2 C_F  \right] \lwc{eu}{RL}[S][prst] \, ,\nn
\nnn
	\dlwc{ed}{RR}[S][prst] &= - \left[ \frac{20}{3} e^2 + 6 g^2 C_F \right] \lwc{ed}{RR}[S][prst] -32 e^2 \lwc{ed}{RR}[T][prst]     \, , \nn
\nnn
	\dlwc{ed}{RR}[T][prst] &= -\frac23 e^2 \lwc{ed}{RR}[S][prst] + \left[ \frac{20}{9} e^2 + 2 g^2 C_F \right] \lwc{ed}{RR}[T][prst]   \, , \nn
\nnn
	\dlwc{ed}{RL}[S][prst] &= - \left[  \frac{20}{3} e^2+ 6 g^2 C_F  \right] \lwc{ed}{RL}[S][prst]  \, , \nn
\nnn
	\dlwc{ee}{RR}[S][prst] &=  16 e^2  \lwc{ee}{RR}[S][ptsr]- 4 e^2  \lwc{ee}{RR}[S][prst] 
	   \, ,
\label{b1}	   
\end{align}
where $\dlwc{}{} = 16\pi^2 \mu\frac{\rd}{\rd \mu} \lwc{}{}$, $C_F = \frac{4}{3}$, and $b_{0,e} = -\frac{4}{3} (n_e + \frac{1}{3} n_d + \frac{4}{3} n_u)$ is the leading coefficient of the QED $\beta$-function.
Note that $\lwc{ee}{RR}[S][prst]=\lwc{ee}{RR}[S][stpr]$ by symmetry of the operator $\op{ee}{S,RR}{}$. The labels $p,r,s,t,w,v$  in Eq.~\eqref{b1} run over the flavors active at a given scale $\mu$.

\bibliographystyle{JHEP}
\bibliography{mue}

\end{document}